\documentclass[11pt]{article}
\def\beq{\begin{equation}}
\def\eeq{\end{equation}}
\def\bea{\begin{eqnarray}}
\def\eea{\end{eqnarray}}

\def\be{\begin{equation}}
\def\bea{\begin{eqnarray}}
\def\ee{\end{equation}}
\def\eea{\end{eqnarray}}

\begin{document}

\begin{flushright}
BRX TH-568 \end{flushright}

\begin{center}
{\large\bf Graviton-Form Invariants in D=11 Supergravity}

{\large S.\ Deser$^1$ and D.\ Seminara$^2$}

{\it $^1$Department of Physics\\ Brandeis University\\
Waltham, MA 02454, USA\\{\tt deser@brandeis.edu}

$^2$Dipartimento di Fisica, Polo Scientifico \\ Universit\`{a} di
Firenze \\ INFN Sezione di Firenze Via G. Sansone 1  \\50019 Sesto
Fiorentino, Italy\\ {\tt seminara@fi.infn.it}}
\end{center}
\noindent {\bf Abstract}: We complete an earlier derivation of the
4-point bosonic scattering amplitudes, and of the corresponding
linearized local supersymmetric invariants in D=11 supergravity,
by displaying the form-curvature, $F^2 R^2$, terms.

Some time ago \cite{sdds}, we presented an efficient method for
constructing explicit bosonic invariants at quartic order in D=11
supergravity. We were stimulated in part by contemporaneous
\cite{bern}   calculations of the lowest, 2-loop order,
divergences of the theory. Our approach to finding counterterms
was first to perform direct tree level calculations of all 4-point
bosonic scattering amplitudes. We then localized these non-local
invariants by removing their denominators, through multiplication
by the Mandelstam variables {\it stu}. These were the
desired-guaranteed (linearly) SUSY-counterterms. Indeed, the
localization enabled us to express them in terms of curvatures and
form field strengths rather than through the original polarization
tensors. The only obstacle we encountered was in the form-graviton
sector, whose explicit ``covariantization" we could not
provide--hence this belated note on a subject that is still of
current \cite{rajaraman} interest.

We will not re-record here the remaining, $R^4 + F^4 + R F^3$,
invariants as they are already given and expounded-on in
\cite{sdds}, where notation and details are found. We emphasize
that both the curvatures $R$ and four-form field strengths $F$ are
on their respective linearized mass shells: Space is Ricci-flat
and $F$ is divergenceless. The (relatively normalized) promised
local $R^2 F^2$ terms are given by the ten combinations
\begin{eqnarray}
      L_{gF}& = & + \frac{1}{24}\;{\cal R}_{\mu_1~\mu_2~\mu_3~\mu_4}\;{\cal R}_{\mu_5~\mu_2~\mu_3~\mu_4}\;{\cal D}_{\mu_1}
     F_{\mu_6~\mu_7~\mu_8~\mu_9}\;{\cal D}_{\mu_5}F_{\mu_6~\mu_7~\mu_8~\mu_9}\nonumber\\
      && - \frac{1}{3}\;{\cal R}_{\mu_1~\mu_2~\mu_3~\mu_4}\;{\cal R}_{\mu_5~\mu_2~\mu_6~\mu_4}\;{\cal D}_{\mu_3}
     F_{\mu_1~\mu_7~\mu_8~\mu_9}\;{\cal D}_{\mu_6}F_{\mu_5~\mu_7~\mu_8~\mu_9}\nonumber\\
      && - \frac{1}{2}\;{\cal R}_{\mu_1~\mu_2~\mu_3~\mu_4}\;{\cal R}_{\mu_5~\mu_2~\mu_6~\mu_4}\;{\cal D}_{\mu_7}
     F_{\mu_1~\mu_5~\mu_8~\mu_9}\;{\cal D}^{\mu_7}F_{\mu_3~\mu_6~\mu_8~\mu_9}\nonumber\\
      && -\frac{2}{3}\;{\cal R}_{\mu_1~\mu_2~\mu_3~\mu_4}\;{\cal R}_{\mu_5~\mu_2~\mu_6~\mu_4}\;F_{\mu_1~\mu_7~
      \mu_8~\mu_9}\;{\cal D}_{\mu_3}{\cal D}_{\mu_6}F_{\mu_5~\mu_7~\mu_8~\mu_9}\nonumber\\
      && + \frac{1}{2}\;{\cal R}_{\mu_1~\mu_2~\mu_3~\mu_4}\;{\cal R}_{\mu_5~\mu_4~\mu_6~\mu_7}\;{\cal D}_{\mu_6}
      F_{\mu_1~\mu_2~\mu_8~\mu_9}\;{\cal D}_{\mu_7}F_{\mu_3~\mu_5~\mu_8~\mu_9}\nonumber\\
      && - \frac{1}{2}\;{\cal R}_{\mu_1~\mu_2~\mu_3~\mu_4}\;{\cal R}_{\mu_5~\mu_4~\mu_6~\mu_7}\;{\cal D}_{\mu_6}
      F_{\mu_3~\mu_5~\mu_8~\mu_9}\;{\cal D}_{\mu_7}F_{\mu_1~\mu_2~\mu_8~\mu_9}\nonumber\\
      && + \frac{1}{6}\;{\cal R}_{\mu_1~\mu_2~\mu_3~\mu_4}\;{\cal R}_{\mu_5~\mu_6~\mu_3~\mu_4}\;{\cal D}_{\mu_5}
     F_{\mu_1~\mu_7~\mu_8~\mu_9}\;{\cal D}_{\mu_6}F_{\mu_2~\mu_7~\mu_8~\mu_9}\nonumber\\
      && + \frac{1}{8}\;{\cal R}_{\mu_1~\mu_2~\mu_3~\mu_4}\;{\cal R}_{\mu_5~\mu_6~\mu_3~\mu_4}\;{\cal D}_{\mu_7}
     F_{\mu_1~\mu_2~\mu_8~\mu_9}\;{\cal D}^{\mu_7}F_{\mu_5~\mu_6~\mu_8~\mu_9}
 \nonumber\\
      && - \frac{1}{2}\;{\cal R}_{\mu_1~\mu_2~\mu_3~\mu_4}\;{\cal R}_{\mu_5~\mu_6~\mu_7~\mu_4}\;{\cal D}_{\mu_8}
      F_{\mu_1~\mu_2~\mu_7~\mu_9}\;{\cal D}^{\mu_8}F_{\mu_5~\mu_6~\mu_3~\mu_9}\nonumber\\
      && - \frac{1}{4}\;{\cal R}_{\mu_1~\mu_2~\mu_3~\mu_4}\;{\cal D}_{\mu_5}F_{\mu_1~\mu_2~\mu_6~\mu_7}
   \;{\cal D}^{\mu_5}{\cal R}_{\mu_8~\mu_9~\mu_3~\mu_4}\;F_{\mu_8~\mu_9~\mu_6~\mu_7} \; .\nonumber
      \end{eqnarray}
We have not seriously attempted to simplify this result using say
cyclic identities, nor to obtain ``current-current"
factorizations; it seems to us unlikely that major condensation
can occur.

\noindent NOTE ADDED: Some time ago, corresponding quartic (in
curvature and 3-form $H$)  invariants were calculated in the
heterotic string slope expansion \cite{gross}. A mutual check
would be to compare them with the $D$=10 KK reduction of our
various invariants, upon identifying $F_{...11}$ with $H_{...}$,
dropping all ${\cal D}_{11}$, and ${\cal R}_{\mu\nu\lambda  11}$.
The famous ``$t_8 t_8$" hallmark of the string expansion seems
likely to emerge here as well.   We are grateful to P.\ Vanhove
for this interesting suggestion.

This work was supported in part by NSF grant PHY04-01667.


\begin{thebibliography}{99}
\bibitem{sdds}
S.\ Deser and D.\ Seminara, Phys.\ Rev.\ {\bf D62}, 084010 (2000);
hep-th/0002241; Phys.\ Rev.\ Lett.\ {\bf 82},  2435 (1999); S.\
Deser, J.S.\ Franklin and D.\ Seminara, Class.\ Quant.\ Grav.\
{\bf 16}, 2815 (1999).

\bibitem{bern}
 Z.\ Bern, L.\ Dixon, D.C.\ Dunbar, M.\ Perelstein and J.S.\
 Rozowsky,
 Nucl.\ Phys.\ {\bf B530}, 401 (1998).

\bibitem{rajaraman}
A.\ Rajaraman, hep-th/0505155.

\bibitem{gross}
D.J.\ Gross and J.H.\ Sloan, Nucl.\ Phys.\  {\bf B291}, 41 (1987).
\end{thebibliography}
\end{document}